\documentclass[cis]{ipart}

\RequirePackage[OT1]{fontenc}
\RequirePackage{amsthm,amsmath,amssymb}
\RequirePackage[numbers,square]{natbib} %
\RequirePackage[colorlinks]{hyperref}
\RequirePackage{graphicx}
\RequirePackage{xcolor}

\newtheorem{thm}{Theorem}

\pubyear{0000}
\volume{0}
\issue{0}
\firstpage{1}
\lastpage{9}

\received{\smonth{12} \sday{7}, \syear{2012}}

\begin{document}

\begin{frontmatter}
\title{A randomized covering-packing duality between source and channel coding\thanksref{T1}}
\thankstext{T1}{Footnote to the title with the `thankstext' command.}

\begin{aug}
\author{\fnms{Mukul} \snm{Agarwal}\thanksref{t1,t2}\ead[label=e1]{magar@alum.mit.edu}},
\address{Department of Electrical and Computer Engineering, \\
University of Toronto\\
\printead{e1}}
\and
\author{\fnms{Sanjoy } \snm{Mitter}\thanksref{t3}\ead[label=e2]{mitter@mit.edu}}
\address{Laboratory for information and decision systems, \\
Department of Electrical Engineering and Computer Science, \\
Massachusetts Institute of Technology\\
\printead{e2}}

\thankstext{t1}{Some comment.}
\thankstext{t2}{First supporter of the project.}
\thankstext{t3}{Second supporter of the project.}
\runauthor{First Author et al.}

\affiliation{Some University and Another University}

\end{aug}

\begin{abstract}

Given a \emph{{  general channel}} $b$ over which the uniform $X$ source, denoted by $U$, is \emph{directly} communicated within distortion $D$. The source $U$ puts uniform distribution on all sequences with type \emph{precisely} $p_X$ as compared with the i.i.d. $X$ source which puts `most of' its mass on sequences with type `close to' $p_X$. A \emph{randomized} covering-packing duality is established between source-coding and channel-coding by considering the source-coding problem (covering problem) of coding the source $U$ within distortion $D$ and the channel coding problem (packing problem) of reliable communication over $b$, thus  leading to a proof of $C \geq R_U(D)$ where $C$ is the capacity of $b$ and $R_U(D)$ is the rate-distortion function of $U$. This  also leads to an operational view of source-channel separation for communication with a fidelity criterion.

\end{abstract}


\end{frontmatter}

\addtolength{\parskip}{0.5\baselineskip}
\setlength{\parindent}{0cm}

\section{Introduction} \label{Introduction}

Given a {  general channel} $b$ over which the uniform $X$ source is \emph{directly} communicated within distortion $D$. 

\begin{figure} 
\begin{center}
\includegraphics[scale = 1.0]{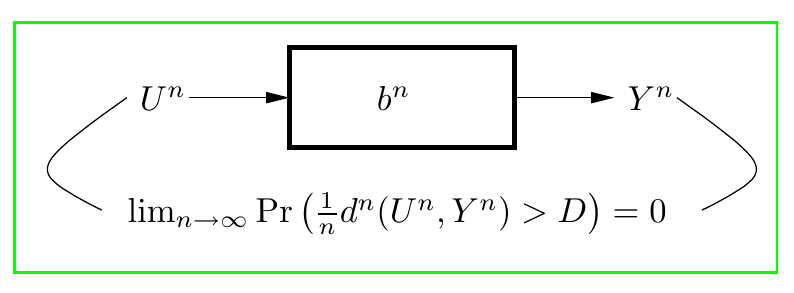}
\caption{A channel which communicates the uniform $X$ source within distortion $D$ }
\label{channelDistortionD}
\end{center}
\end{figure}

This means the following: 

Let the source input space be $\mathcal X$ and the source reproduction space be $\mathcal Y$. { $\mathcal X$ and $\mathcal Y$ are finite sets.}  Intuitively, a uniform $X$ source, $U$, puts a uniform distribution on all sequences with a type $p_X$. This is as opposed to the i.i.d. $X$ source which puts ``most of'' its mass on sequences with type ``close to'' $p_X$. See Section \ref{NotAndDefIII} for a precise definition. A {  general channel} is a sequence $<b^n>_1^\infty$ where $b^n$ is a transition probability from $\mathcal X^n$ to $\mathcal Y^n$; a precise definition of a {  general channel} can be found in Section \ref{NotAndDefIII}. When the block-length is $n$, the uniform $X$ source is denoted by $U^n$. With input $U^n$ into the channel, the output is $Y^n$, and is such that 
\begin{align}
\lim_{n \to \infty} \Pr \left ( \frac{1}{n} d^{n}(U^{n}, Y^{n}) > D \right ) = 0
\end{align}
{ where $<d^n>_1^\infty$, $d^n: \mathcal X^n \times \mathcal Y^n \to [0, \infty )$, is a permutation-invariant (a special case is additive) distortion function.}  { The generality of the channel is in the sense of Verdu and Han \cite{VerduHan}. See Section \ref{NotAndDefIII}  for precise definitions.}  {See Figure \ref{channelDistortionD}}.

\begin{figure} 
\begin{center}
\includegraphics[scale = 1.0]{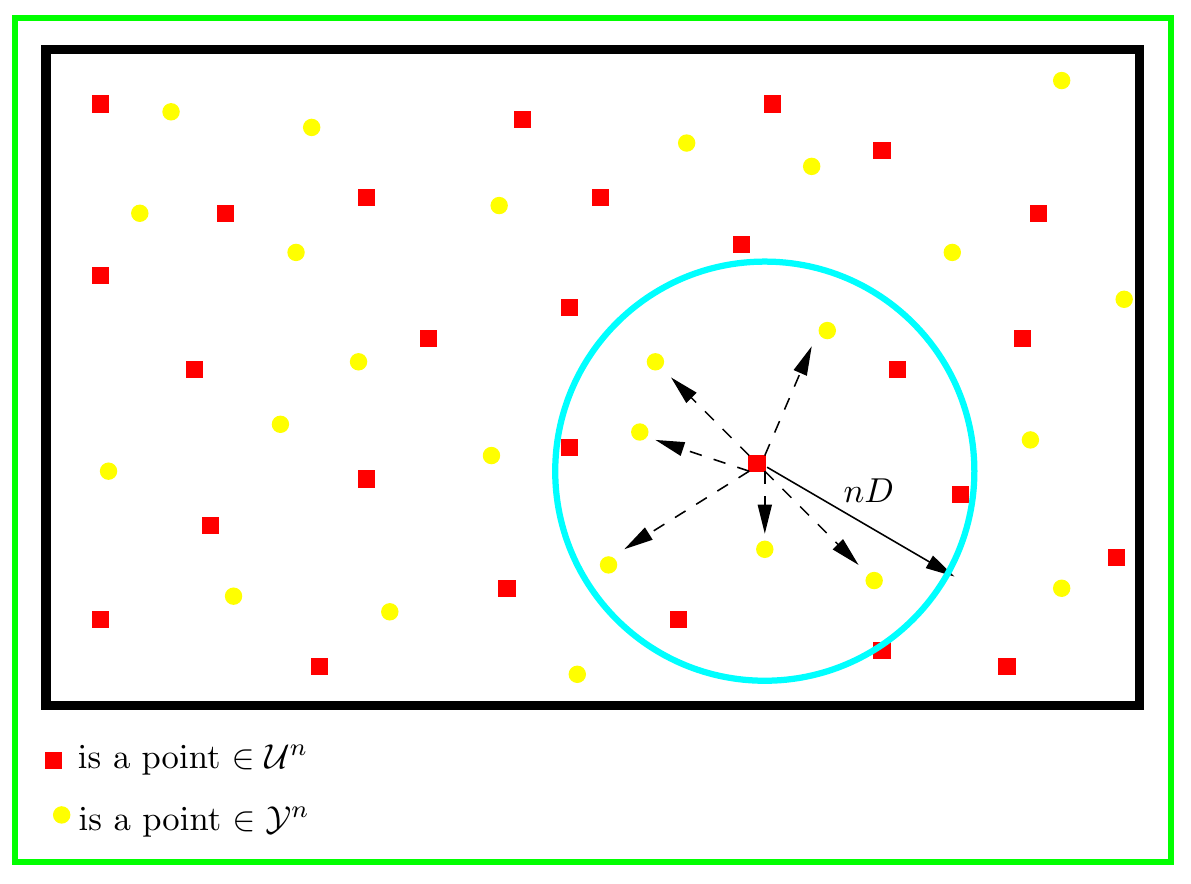}
\caption{Intuitive action of a channel which directly communicates the uniform $X$ source within a distortion $D$ }
\label{intuitiveChannel}
\end{center}
\end{figure}

Such a {  general channel} intuitively functions as follows: when the block-length is $n$, with high probability, a sequence in $u^n \in \mathcal U^{n}$ is distorted within a ball of radius $nD$  and this probability $\to 0$ as $n \to \infty$. Note that $u^n \in \mathcal U^n$ but the ball of radius $nD$ exists in the output space $\mathcal Y^n$. { See Figure \ref{intuitiveChannel}}.

{
\emph{Note that the uniform $X$ source is not defined for all block-lengths; this point will be clarified in Section \ref{NotAndDefIII}}.}

Consider the two problems:
\begin{itemize}
\item
Covering problem: the rate-distortion source-coding problem of compressing the source $U$ within distortion $D$, that is, computing the minimum rate needed to compress the source $U$ within a distortion $D$. Denote the rate-distortion function by $R_U(D)$. Intuitively, the question is to find the minimum number of $y^n \in \mathcal Y^n$ such that balls of radii $nD$ circled around $y^n$ \emph{cover} the space $\mathcal U^n$. Note that balls are circled on $y^n \in \mathcal Y^n$ but balls of radius $nD$ exist in $\mathcal U^n$. Since the setting is information-theoretic, the balls should `almost' cover the whole space.
\item
Packing problem: the channel-coding problem of communicating reliably over a {  general channel} $b$ which is known to directly communicate the source $U$ within distortion $D$ (packing problem). Denote the channel capacity by $C$. Intuitlvely, the question is to find the maximum number of $u^{n} \in \mathcal U^n$ such that balls of radii $nD$ circled around these $u^n$ \emph{pack} the $\mathcal Y^n$ space. Note that $u^n \in \mathcal U^n$ but balls of radil $nD$ circled around these codewords exist in the $\mathcal Y^n$ space. Since the setting is information theoretic, the balls which pack the space can overlap `a bit'.
\end{itemize}

Clearly, there is a duality in these problem statements. It is unclear how to make this duality precise for these deterministic problems. However, a randomized covering-packing duality can be established between the above two problems, thus also proving that the answer to the first problem is less than or equal to the answer to the second problem, in the following way:

The codebook construction and error analysis for the source-coding problem are roughly the following: 
Let the block-length be $n$. Generate $2^{nR}$ codewords $\in \mathcal Y^n$ independently and uniformly from the set of all sequences with type $q$ where $q$ is an achievable type on the output space. \emph{Roughly}, a $u^n \in \mathcal U^n$ is encoded via minimum distance encoding. The main error analysis which needs to be carried out is the probability that a certain codebook sequence does not encode a particular $u^n$, that is,
\begin{align}\label{E1}
\Pr \left ( \frac{1}{n} d^n(u^n, Y^n) > D \right )
\end{align}
where $Y^n$ is a uniform random variable on sequences of type $q$. A best possible $q$ is chosen in order to get an upper bound on the rate-distortion function.

The codebook construction and error analysis for the channel-coding problem are roughly the following: 
Let the block length be $n$. Generate $2^{nR}$ codewords $\in \mathcal U^n$ independently and uniformly. Let $y^n$ be received. The decoding of which codeword is transmitted is \emph{roughly} via minimum distance decoding. As will become clearer later, the main error calculation in the channel-coding problem is the probability of correct decoding for which the following needs to be calculated:
\begin{align}\label{E2}
\Pr \left ( \frac{1}{n} d^n(U^n, y^n) > D \right )
\end{align}
where $y^n$ has type $q$. Finally, a worst case error analysis is done by taking the worst possible $q$.

By symmetry, (\ref{E1}) and (\ref{E2}) are equal assuming the distortion function is additive (more generally, permutation invariant) and this leads to a proof that $C \geq R_U(D)$. The above steps will be discussed in much detail, later in this paper. \emph{This equality of (\ref{E1}) and (\ref{E2})  is a randomized covering-packing connection, and is a duality between source-coding and channel-coding.} Further, this is an operational view and proof in the sense that only the operational meanings of channel capacity as the maximum rate of reliable communication and the rate-distortion function as the minimum rate needed to compress a source with certain distortion are used. Of  course, certain randomized codebook constructions are used. No functional simplifications beyond the equality of (\ref{E1}) and (\ref{E2}) are needed.

{ This proof is discussed precisely in Section \ref{RCPD} and intuitively in Appendix \ref{AppendixIntuitive}.}

 If $b$ is the composition of an encoder, channel and decoder, that is, $b^n = e^n \circ k \circ f^n$ for some encoder, decoder, $<e^n, f^n>_1^\infty$ and channel $k$ and the uniform $X$ source is communicated over this channel by use of some encoder-decoder $<E^n, F^n>_1^\infty$. Then, it follows that by use of encoder-decoder $<E^n \circ e^n, f^n \circ F^n>_1^\infty$, reliable communication can be accomplished over channel $k$ at rates $<R_U(D)$. By use of the argument of source-coding followed by channel-coding, optimality of source-channel separation for communication of the uniform $X$ source over the channel $k$. \emph{This leads to an operational view of source-channel separation for communication with a fidelity criterion.} Note that both the channel capacity problem and the rate-distortion problem are infinite dimensional optimization problems. By use of this methodology, the optimality of source-channel separation is proved without reducing the problems to finite dimensional problems. This is as opposed to the proof of separation, for example, in \cite{Shannon} which crucially relies on the the single-letter maximum mutual information expression for channel capacity and the single-letter minimum mutual information expression for the rate-distortion function.

{  Since the decoding rule for the channel-coding problem depends only on the end-to-end description that the channel communicates the uniform $X$ source within distortion $D$, in addition to a general channel, assuming random codes are permitted, duality also holds for a compound channel, that is, where the channel belongs to a set, (see for example \cite{CsiszarKorner} for a discussion on compound channels). Note that the channel model is still general. For the same reason, source-channel separation for communication with a fidelity criterion also holds for a general, compound channel assuming random codes are permitted. This will be discussed in some detail, later}.

\emph{An operational view, as regards this paper, refers to  a view which uses only the operational meanings of quantities: for example, of channel capacity as the maximum rate of reliable communication or the rate-distortion function as the minimum rate needed to code a source with a certain distortion} It does \emph{not} mean constructive.

The source $U$ is ideal for this purpose because it puts mass only on the set of sequences with a particular type. If one tries to carry out the above argument for the i.i.d. $X$ source, $\epsilon$s and $\delta$s enter the picture. A generalization to the i.i.d. $X$ source can be made via a perturbation argument.

\section{Literature Survey}

Duality between source-coding and channel-coding has been discussed in a number of settings in the information-theory literature.

Shannon \cite{Shannon} discussed, on a high level,  a functional duality between source-coding and channel-coding by considering a channel-coding problem where there is a cost associated with different input letters which amounts to finding a source which is just right for the channel and desired cost. Similarly, the rate-distortion source-coding problem corresponds to finding a channel that is just right for the source and the allowed distortion level.  Further, Shannon makes the statement, ``This duality can be pursued further and is related to a duality between past and future and notions of control and knowledge. Thus we may have knowledge of the past but cannot control it; we may control the future but have no knowledge of it.'' 

A general formulation of this functional duality has been posed in \cite{Ramchandran} which considers the channel capacity with cost constraints problem and the rate-distortion problem, defines when the problems are duals of each other, and proves that channel capacity is equal to the rate-distortion function if the problems are dual. The purpose of our paper is not a functional duality or a mathematical programming based duality, but a operational duality where operational is defined in the previous section.

Operational duality, as defined by Ankit et al \cite{Ankit} refers to the property that optimal encoding/decoding schemes for one problem lead to optimal encoding/decoding schemes for the corresponding dual problem. They show that if used as a lossy compressor, the maximum-likelihood channel decoder of a randomly chosen capacity-achieving codebook achieves the rate-distortion function almost surely . \emph{Note that the definition of operational used in \cite{Ankit} is different from the definition of operational used in this paper.}

Csiszar and Korner \cite{CsiszarKorner} prove the rate-distortion theorem by first constructing a ``backward'' DMC and codes for this DMC such that source-codes meeting the distortion criterion are obtained from this channel code by using the channel decoder as a source encoder and vice-versa; for this purpose, channel codes with large error probability are needed. The view-point is suggestive of a duality between source and channel coding. There is no backward channel in our paper: there is a forward channel which directly communicates the source $U$ within distortion $D$ and there is the rate-distortion source-coding problem.

{  Yassaee \cite{Yassaee} have studied duality between channel coding problem and secret-key agreement problem (in the source-model sense) They show how an achievability proof for each of these problems can be converted into an achievability proof for the other one. }

{The decoding rule used in this paper is a variant of a minimum distance decoding rule. For discrete memoryless channels, decoders minimizing a distortion measure have been studied as mis-match decoding and are suboptimal in general though optimal if the distortion measure is matched, that is, equal to the negative log of the channel transition probability; see for example the paper of Csiszar and Narayan \cite{CsiszarNarayan}.}

The results in this paper form a part of the first authors Ph. D. dissertation \cite{MukulPhdThesis}.

{  Recall the important point that the duality between source-coding and channel-coding, as discussed  in this paper is operational in the sense it uses only the operational meanings of channel capacity as the maximum rate of reliable communications and the rate-distortion function as the minimum rate needed to code a source with certain distortion levels, and this sense is different from the sense in which duality is discussed in the above mentioned papers. Major functional simplifications are not used. Random codes are constructed for both problems and a connection is seen between the two problems, which leads to a randomized covering-packing duality. }

\section{Notation and definitions} \label{NotAndDefIII}

Superscript $n$ will denote a quantity related to block-length $n$. For example, $x^n$ will be the channel input when the block-length is $n$. As block-length varies, $x = <x^n>_1^\infty$ will denote the sequence for various block-lengths.

The source input space is $\mathcal X$ and the source reproduction space is $\mathcal Y$. $\mathcal X$ and $\mathcal Y$ are finite sets. $X$ is a random variable on $\mathcal X$. Let $p_X(x)$ be rational $\forall x$. Let $n_0$ denote the least positive integer for which $n_0p_X(x)$ is an integer $\forall x \in \mathcal X$. Let $\mathcal U^n$ denote the set of sequences with (\emph{exact}) type $p_X$. $\mathcal U^n$ is non-empty if and only if $n_0$ divides $n$. Let $n' \triangleq n_0n$. Let $U^{n'}$ denote a random variable which is uniform on $\mathcal U^{n'}$ and zero elsewhere. Then, $<U^{n'}>_1^\infty$ is the uniform $X$ source and is denoted by $U$. The uniform $X$ source can be defined only for those $X$ for which $p_X(x)$ is rational $\forall x \in \mathcal X$.

{ 
\emph{Every mathematical entity which had a superscript $n$ in Section \ref{Introduction}  will have a superscript $n'$ henceforth. This is because the uniform $X$ source is defined only for block-lengths $n'$. The reader is urged not to get confused between this change of superscript between Section \ref{Introduction} and the rest of this paper. Further, the reader is urged to read Section \ref{Introduction} by replacing $n$ with $n'$ in mathematical entities.} 
}

Let $q$ denote a type on the set $\mathcal Y$ which is achievable when the block-length is $n'$. $\mathcal V_q^{n'}$ is the set of all sequences with type $q$. The uniform distribution on $\mathcal V_q^{n'}$ is $V_q^{n'}$.

Since the uniform $X$ source is defined only for block-lengths $n'$, distortion function, channels, encoders and decoders will be defined only for block-lengths $n'$.

$d = <d^{n'}>_1^\infty$ is the distortion function where $d^{n'}: \mathcal X^{n'} \times \mathcal Y^{n'} \rightarrow [0, \infty )$. Let $\pi^{n'}$ be a permutation (rearrangement) of $(1, 2, \ldots, n')$. That is, for $1 \leq i \leq n'$, $\pi^{n'}(i) \in \{1, 2, \ldots, n' \}$ and that, $\pi^{n'}(i)$, $1 \leq i \leq n'$ are different. For $x^{n'} \in \mathcal X^{n'}$, denote 
\begin{align}
\pi^{n'}x^{n'} \triangleq (x^{n'}(\pi^{n'}(1)), x^{n'}(\pi^{n'}(2)), \ldots, x^{n'}(\pi^{n'}(n')))
\end{align}
For $y^{n'} \in \mathcal Y^{n'}$, $\pi^{n'}y^{n'}$ is defined analogously. $<d^{n'}>_1^\infty$ is said to be permutation invariant if $\forall n'$,
\begin{align}
d^{n'}(\pi^{n'}x^{n'}, \pi^{n'}y^{n'}) = d^{n'}(x^{n'}, y^{n'}), \forall x^{n'}\in \mathcal X^{n'}, y^{n'} \in \mathcal Y^{n'} 
\end{align}
An additive distortion function is defined as follows. Let $d: \mathcal X \times \mathcal Y \rightarrow [0, \infty)$ be a function. Define
\begin{align}
d^{n'}(x^{n'}, y^{n'}) = \sum_{i=1}^{n'} d(x^{n'}(i), y^{n'}(i))
\end{align}
Then, $<d^{n'}>_1^\infty$ is an additive distortion function.

Additive distortion functions are special cases of permutation invariant distortion function. Except at the end of the paper where conditions are derived for a certain technical conditions to be true for which additive distortion functions will be required, most of this paper will use permutation invariant distortion functions.

A {  general channel} $b = <b^{n'}>_1^\infty$ is defined as follows:

The input space of the channel is $\mathcal X$ and the output space is $\mathcal Y$. 
\begin{align}
b^{n'}:&\mathcal X^{n'} \rightarrow \mathcal P(\mathcal Y^{n'}) \\ 
          & x^{n'}                  \rightarrow b^{n'}(y^{n'}|x^{n'}) \nonumber
 \end{align}
$b^{n'}(y^{n'}|x^{n'})$ should be thought of as the probability that the output of the channel is $y^{n'}$ given that the input is $x^{n'}$.

{  Note that the channel model is general in the sense of Verdu and Han \cite{VerduHan}.  }

Let 
\begin{align}
\mathcal M^{n'}_R \triangleq \{1, 2, \ldots, 2^{\lfloor n'R \rfloor} \}
\end{align}
$\mathcal M^{n'}_R$ is the message set. When the block-length is $n'$, a rate $R$ deterministic source encoder  is $e_s^{n'}: \mathcal X^{n'} \rightarrow \mathcal M_R^{n'}$ and a rate $R$ deterministic source decoder $f_s^{n'}: \mathcal M_R^{n'} \rightarrow \mathcal Y^{n'}$.  $(e_s^{n'}, f_s^{n'})$ is the block-length $n'$ rate $R$  deterministic source-code. The source-code is allowed to be random in the sense that encoder-decoder is a joint probability distribution on the space of deterministic encoders and decoders. $<e_s^{n'}, f_s^{n'}>_1^\infty$ is the rate $R$ source-code. The classic argument used in \cite{Shannon} to prove the achievability part of the rate-distortion theorem uses a random source code.

When the block-length is $n'$, a rate $R$ deterministic channel encoder is a map $e_c^{n'}:\mathcal M_R^{n'} \rightarrow \mathcal X^{n'}$ and a rate $R$ deterministic channel decoder is a map $f_c^{n'}: \mathcal Y^{n'} \rightarrow \hat {\mathcal M}_R^{n'}$ where $\hat {\mathcal M}_R^{n'} \triangleq \mathcal M_R^{n'} \cup \{e\}$ is the message reproduction set where `e' denotes error. The encoder and decoder are allowed to be random in the sense discussed previously. $<e_c^{n'}, f_c^{n'}>_1^\infty$ is the rate $R$ channel code. The classic argument used in \cite{ShannonReliable} to derive the achievability of the mutual information expression for channel capacity uses a random channel code.

The source-code $<e_s^{n'}, f_s^{n'}>_1^\infty$ is said to code the source $U$ to within a distortion $D$ if with input $U^{n'}$ to $e_s^{n'} \circ f_s^{n'}$, the output is $Y^{n'}$ such that
\begin{align} \label{SourceCodeD}
\lim_{n' \to \infty} \Pr \left ( \frac{1}{n'} d^{n'}(U^{n'}, Y^{n'}) > D \right ) = 0
\end{align}
(\ref{SourceCodeD}) is the probability of excess distortion criterion.
The infimum of rates needed to code the uniform $X$ source to within the distortion $D$ is the rate-distortion function $R^P_U(D)$. If $\lim$ in (\ref{SourceCodeD}) is replaced with $\lim \inf$, the criterion is called the $\inf$ probability of excess distortion criterion and the corresponding rate-distortion function is denoted by $R^P_U(D, \inf)$.

Denote 
\begin{align}
g = <g^{n'}>_1^\infty  \triangleq <e_c^{n'} \circ b^{n'} \circ f_c^{n'}>_1^\infty
\end{align}
Then, $g$ is a general channel with input space $\mathcal M_R^{n'}$ and output space $\hat {\mathcal M}_R^{n'}$. Rate $R$ is said to be reliably achievable over $b$ if there exists a  rate $R$ channel code $<e_c^{n'}, f_c^{n'}>_1^\infty$ such that 
\begin{align}
\lim_{n' \to \infty}  \sup_{m^{n'} \in \mathcal M_R^{n'}} g^{n'}(\{ m^{n'}\}^c|m^{n'}) = 0
\end{align}
Supremum of all achievable rates is the capacity of $b$.

The channel $b$ is said to communicate the source $U$ \emph{directly} within distortion $D$ if with input $U^{n'}$ to $b^{n'}$, the output is $Y^{n'}$ such that
\begin{align} 
\lim_{n' \to \infty} \Pr \left ( \frac{1}{n'} d^{n'}(U^{n'}, Y^{n'}) > D \right ) = 0
\end{align}
See Figure \ref{channelDistortionD} in Section \ref{Introduction} with $n$ replaced by $n'$.

{ 
In this paper, only the end-to-end description of a channel $<{b^{n'}}_1^\infty$ which communicates the uniform $X$ source directly within distortion $D$ is used and not the particular $b^{n'}$; for this reason, the {  general channel} should be thought of as a \emph{black-box} which communicates the uniform $X$ source within distortion $D$.}

In order to draw the randomized covering-packing duality between source and channel coding, the source-coding problem which will be considered is that of coding the source $U$ within distortion $D$ and the channel coding problem which will be considered is the rates of reliable communication over a $b$ which communicates the source $U$ directly within distortion $D$. A relation will be drawn between the rate-distortion function for the uniform $X$ source and the capacity of $b$ and in the process, the randomized covering-packing duality will emerge.

\section{Randomized covering-packing duality}  \label{RCPD}

\begin{thm} \label{KKUniUniKK}
Let $b$ directly communicate source $U$ within distortion $D$ under a permutation invariant distortion function $d$. Assume that $R^P_U(D) = R^P_U(D, \inf)$. Then, reliable communication can be accomplished over $b$ at rates $<R^P_U(D)$. In other words, the capacity of $b$, $C \geq R^P_U(D)$.
\end{thm}

Note that the technical condition $R^P_U(D) = R^P_U(D, \inf)$ can be proved for an additive distortion function. See the discussion following the proof of the theorem.

\begin{proof}
This will be done by use of parallel random-coding arguments for two problems:
\begin{itemize}
\item
\emph{Channel-coding problem:} 
Rates of reliable communication over $b$.
\item
\emph{Source-coding problem:} 
Rates of coding  for the uniform $X$ source with a distortion $D$ under the $\inf$ probability of excess distortion criterion.
\end{itemize}

\emph{Codebook generation:}
\begin{itemize}
\item
\emph{Codebook generation for the channel-coding problem:}
Let reliable communication be desired at rate $R$. Generate $2^{\lfloor n'R \rfloor}$ sequences independently and uniformly from $\mathcal U^{n'}$. This is the codebook $\mathcal K^{n'}$.
\item
\emph{Codebook generation for the source-coding problem:}
Let source-coding be desired at rate $R$. Generate $2^{\lfloor n'R \rfloor}$ codewords  independently and uniformly from $\mathcal V_q^{n'}$ for some type $q$ on $\mathcal Y$ which is achievable for block-length $n'$. This is the codebook $\mathcal L^{n'}$. 
\end{itemize}

\emph{Joint typicality:}

Joint typicality for both the channel-coding and source-coding problems is defined as follows:
$(u^{n'}, y^{n'}) \in \mathcal U^{n'} \times \mathcal Y^{n'}$ jointly typical if 
\begin{align}
\frac{1}{n'} d^{n'} (u^{n'}, y^{n'}) \leq D
\end{align}

\emph{Decoding and encoding:}
\begin{itemize}
\item
\emph{Decoding for the channel-coding problem:}
Let $y^{n'}$ be received. If there exists unique $u^{n'} \in \mathcal K^{n'}$ for which $(u^{n'}, y^{n'})$ jointly typical, declare that $u^{n'}$ is transmitted, else declare error. 
\item
\emph{Encoding for the source-coding problem:}
Let $u^{n'} \in \mathcal U^{n'}$ need to be source-coded. If there exists some $y^{n'} \in \mathcal L^{n'}$, encode $u^{n'}$ to one such $y^{n'}$, else declare error.
\end{itemize}

\emph{Some notation:}
\begin{itemize}
\item
\emph{Notation for the channel-coding problem:} 
Let message $m^{n'} \in \mathcal M_R^{n'}$ be transmitted. Codeword corresponding to $m^{n'}$ is $u_c^{n'}$. Non-transmitted  codewords are ${u'}_1^{n'}, {u'}_2^{n'}, \ldots, {u'}_{2^{\lfloor n'R \rfloor} - 1}^{n'}$. $u_c^{n'}$ is a realization of $U_c^{n'}$. $U_c^{n'}$  is uniform on $\mathcal U^{n'}$. ${u'}_i^{n'}$ is a realization of ${U'}_i^{n'}$. ${U'}_i^{n'}$ is uniform on $\mathcal U^{n'}$, $1 \leq i \leq 2^{\lfloor n'R \rfloor} - 1$. $U_c^{n'}, {U'_i}^{n'}, 1 \leq i \leq 2^{\lfloor n'R \rfloor} - 1$ are independent of each other. The channel output is $y^{n'}$. $y^{n'}$ is a realization of $Y^{n'}$. $y^{n'}$ may depend on $u_c^{n'}$ but does not depend on ${u'_i}^{n'}, 1 \leq i \leq 2^{\lfloor n'R \rfloor} - 1$.  As random variables, $Y^{n'}$ and $U_c^{n'}$ might be dependent but $Y^{n'}, {U'_i}^{n'}, 1 \leq i \leq 2^{\lfloor n'R \rfloor} - 1$ are independent. If the type $q$ of the sequence $y^{n'}$ needs to be explicitly denoted, the sequence is denoted by $y_q^{n'}$. $\mathcal G^{n'}$ is the set of all achievable types $q$ on $\mathcal Y$ for block-length $n'$. 

\item
\emph{Notation for the source-coding problem:}
$u_s^{n'}$ is the sequence which needs to be source-coded. $u_s^{n'}$ is a realization of $U_s^{n'}$ which is uniformly distributed on $\mathcal U^{n'}$. The codewords are $y_{q,i}^{n'}, 1 \leq i \leq 2^{\lfloor n'R \rfloor}$ where $q$ denotes the type. $y_{q, i}^{n'}$ is a realization of $V_{q, i}^{n'}, 1 \leq i \leq 2^{\lfloor n'R \rfloor}$ where $V_{q, i}^{n'}$ is uniformly distributed on the subset of $\mathcal Y^{n'}$ consisting of all sequences with type $q$. $u_s^{n'}, y_{q, i}^{n'}, 1 \leq i \leq 2^{\lfloor n'R \rfloor}$ are independently generated; as random variables, $U_s^{n'}, Y_{q, i}^{n'}, 1 \leq i \leq 2^{\lfloor n'R \rfloor}$ are independent. $\mathcal G^{n'}$ is the set of all achievable types $q$ on $\mathcal Y$ for block-length $n'$
\end{itemize}

\emph{Error analysis:}
For the channel-coding problem, the probability of correct decoding is analyzed and for the source-coding problem, the probability of error is analyzed.
\begin{itemize}
\item
\emph{Error analysis for the channel-coding problem:}
From the encoding-decoding rule, it follows that the event of correct decoding given that a particular message is transmitted is 
\begin{align}\label{CorrectDecodingEvent}
& \left \{ \frac{1}{n'} d^{n'}(U_c^{n'}, Y^{n'}) \leq D \right \} \cap
 \cap_{i = 1}^{2^{\lfloor n'R \rfloor} - 1} \left \{ \frac{1}{n'} d^{n'} ({U'}_i^{n'}, Y^{n'}) > D  \right \}
\end{align}
\item
\emph{Error analysis for the source-coding problem:}
From the encoding-decoding rule, it follows that the error event given that a particular message needs to be source-coded is
\begin{align}\label{ErrorEvent}
\cap_{i=1}^{2^{\lfloor n'R \rfloor}} \left \{  \frac{1}{n'}d^{n'}(u^{n'}, V_{q, i}^{n'}) > D \right \}
\end{align}
\end{itemize}
Note that there is choice of $q$ for codebook generation.

\emph{Calculation:}
\begin{itemize}
\item
\emph{Calculation of the probability of correct decoding for the channel-coding problem:}


Bound for probability of event (\ref{CorrectDecodingEvent}):

\begin{align}
       & \Pr \left ( \left \{ \frac{1}{n'} d^{n'}(U_c^{n'}, Y^{n'}) \leq D \right \}  \cap 
        \cap_{i = 1}^{2^{\lfloor n'R \rfloor} - 1} \left \{ \frac{1}{n'} d^{n'} ({U'}_i^{n'}, Y^{n'}) > D  \right \} \right ) \\ 
=    & \Pr \left ( \left \{ \frac{1}{n'} d^{n'}(U_c^{n'}, Y^{n'}) \leq D \right \} \right ) + 
 \Pr \left ( \cap_{i = 1}^{2^{\lfloor n'R \rfloor} - 1} \left \{ \frac{1}{n'} d^{n'} ({U'}_i^{n'}, Y^{n'}) > D  \right \}\right ) -  \nonumber \\
       &  \hspace{2cm} \Pr \left ( \left \{ \frac{1}{n'} d^{n'}(U_c^{n'}, Y^{n'}) \leq D \right \}  \cup 
        \cap_{i = 1}^{2^{\lfloor n'R \rfloor} - 1} \left \{ \frac{1}{n'} d^{n'} ({U'}_i^{n'}, Y^{n'}) > D  \right \} \right ) \nonumber \\
\geq & (1 - \omega_{n'}) + 
\Pr \left ( \cap_{i = 1}^{2^{\lfloor n'R \rfloor} - 1} \left \{ \frac{1}{n'} d^{n'} ({U'}_i^{n'}, Y^{n'}) > D  \right \}\right ) - 1 \nonumber \\
=       & - \omega_{n'} + \Pr \left ( \cap_{i = 1}^{2^{\lfloor n'R \rfloor} - 1} \left \{ \frac{1}{n'} d^{n'} ({U'}_i^{n'}, Y^{n'}) > D  \right \}\right ) \nonumber \\
=       & - \omega_{n'} + \prod_{i=1}^{2^{\lfloor n'R \rfloor } - 1} \Pr \left ( \left \{ \frac{1}{n'} d^{n'} ({U'}_i^{n'}, Y^{n'}) > D \right \} \right ) \nonumber \\ 
          & \hspace{1cm} \mbox{ (since ${U'}_i^{n'}, 1 \leq i \leq 2^{\lfloor n'R \rfloor } - 1$, $Y^{n'}$ are  independent random variables)} \nonumber \\
=       & - \omega_{n'}  +  \left [ \Pr \left ( \left \{ \frac{1}{n'} d^{n'} ({U}^{n'}, Y^{n'}) > D \right \} \right )  \right ] ^ { 2^{\lfloor n'R \rfloor } - 1} \nonumber \\
         & \hspace{1cm} \mbox{(where $U^{n'}$ is uniform on $\mathcal U^{n'}$ and is independent of $Y^{n'}$)} \nonumber
\end{align}
\begin{align}
\nonumber =       &  - \omega_{n'} +   \left [   \sum_{y^{n'} \in \mathcal Y^{n'}} p_{Y^{n'}}(y^{n'}) \Pr \left ( \frac{1}{n'} d^{n'} ({U}^{n'}, Y^{n'})  > D \ \Bigg | \ Y^{n'} = y^{n'}  \right )      \right ] ^ { 2^{\lfloor n'R \rfloor } - 1} \nonumber \\
=       &  - \omega_{n'} +  \left [   \sum_{y^{n'} \in \mathcal Y^{n'}} p_{Y^{n'}}(y^{n'}) \Pr \left ( \frac{1}{n'} d^{n'} ({U}^{n'}, y^{n'})  > D \ \Bigg | \ Y^{n'} = y^{n'}  \right )      \right ] ^ { 2^{\lfloor n'R \rfloor } - 1} \nonumber \\
=       &  - \omega_{n'} +    \left [   \sum_{y^{n'} \in \mathcal Y^{n'}} p_{Y^{n'}}(y^{n'}) \Pr \left ( \frac{1}{n'} d^{n'} ({U}^{n'}, y^{n'})   > D  \right )      \right ] ^ { 2^{\lfloor n'R \rfloor } - 1} \nonumber \\
                   & \hspace{1cm} \mbox{ (since $U^{n'}$ and $Y^{n'}$ are independent) } \nonumber \\
\geq    &  -\omega_{n'} +   \left [ \inf_{y^{n'} \in \mathcal Y^{n'}} \Pr \left ( \left \{ \frac{1}{n'} d^{n'} ({U}^{n'}, y^{n'})   > D \right \} \right ) \right ] ^{ 2^{\lfloor n'R \rfloor } - 1} \nonumber \\
=          &  -\omega_{n'} +  \left [ \inf_{q \in \mathcal G^{n'}} \Pr \left ( \left \{ \frac{1}{n'} d^{n'} ({U}^{n'}, y_q^{n'})   > D \right \} \right ) \right ] ^{ 2^{\lfloor n'R \rfloor } - 1} \nonumber
\end{align}
The last equality above follows because
\begin{align}
\Pr \left ( \left \{ \frac{1}{n'} d^{n'} ({U}^{n'}, y^{n'}) > D \right \} \right )
\end{align}
depends only on the type of $y^{n'}$; see the symmetry argument later.


Rate $R$ is achievable if 
\begin{align}
 &  -\omega_{n'} +  
         \left [ \inf_{q \in \mathcal G^{n'}} \Pr \left ( \left \{ \frac{1}{n'} d^{n'} ({U}^{n'}, y_q^{n'})  
         > D \right \} \right ) \right ] ^{ 2^{\lfloor n'R \rfloor } - 1} 
                   \to 1 \ \mbox{as} \ n' \to \infty
\end{align}
Since $\omega_{n'} \to 0$ as $n' \to \infty$, rate $R$ is achievable if 
\begin{align}\label{JCorrectCalculationJ}
          & \left [ \inf_{q \in \mathcal G^{n'}} \Pr \left ( \left \{ \frac{1}{n'} d^{n'} ({U}^{n'}, y_q^{n'}) > D \right \} \right ) \right ] ^{ 2^{\lfloor n'R \rfloor } - 1} 
           \to 1 \ \mbox{as} \ n' \to \infty
\end{align}
\item
\emph{Calculation of probability of error for the source-coding problem:}

Bound for probability of event (\ref{ErrorEvent}) is calculated using standard arguments:
\begin{align}
           & \Pr \left ( \cap_{i=1}^{2^{\lfloor n'R \rfloor}} \left \{  \frac{1}{n'}d^{n'}(u^{n'}, V_{q, i}^{n'}) > D \right \} \right )  \\ 
          = &   \prod_{i=1}^{2^{\lfloor n'R \rfloor}}  \Pr \left (  \left \{  \frac{1}{n'}d^{n'}(u^{n'}, V_{q, i}^{n'}) > D \right \} \right )   \nonumber \\ 
        = &  \left [  \Pr \left (  \left \{  \frac{1}{n'}d^{n'}(u^{n'}, V_{q, i}^{n'}) > D \right \} \right )\right ] ^{2^{\lfloor n'R \rfloor }} \nonumber
\end{align}
where $V_q^{n'}$ is uniform on $\mathcal V_q^{n'}$.

There is choice of $q \in \mathcal G^{n'}$. Thus, a bound for the probability of the event is
\begin{align}
 \left [  \inf_{q \in \mathcal G^{n'}} \Pr \left (  \left \{  \frac{1}{n'}d^{n'}(u^{n'}, V_q^{n'}) > D \right \} \right )\right ] ^{2^{\lfloor n'R \rfloor }}
\end{align}


Since the $\inf$ probability of excess distortion criterion is used, it follows that rate $R$ is achievable if 
\begin{align}\label{JSourceErrorJ}
\left [  \inf_{q \in \mathcal G^{n'_i}} \Pr \left (  \left \{  \frac{1}{n'_i}d^{n'_i}(u^{n'_i}, V_q^{n'_i}) > D \right \} \right )\right ] ^{2^{\lfloor n'_iR \rfloor }}  \to 0 \ \mbox{for some} \ n'_i = n_0 n_i, \ n_i \  \to \infty
\end{align}
\end{itemize}

\emph{Connection/Duality between channel-coding and source-coding:}

The calculation required in the channel-coding problem is 
\begin{align} \label{CCCalc}
\inf_{q \in \mathcal G^{n'}}\Pr \left ( \left \{ \frac{1}{n'} d^{n'} ({U}^{n'}, y_q^{n'}) > D \right \} \right )
\end{align}
and the calculation required in the source-coding problem is
\begin{align} \label{SCCalc}
\inf_{q \in \mathcal G^{n'}} \Pr \left ( \left \{ \frac{1}{n'} d^{n'} ({u}^{n'}, V_q^{n'}) > D \right \} \right )
\end{align}
It will be proved that (\ref{CCCalc}) and (\ref{SCCalc}) are equal. It will be proved more generally that 
\begin{align}\label{JJMainDualityJJ}
& \Pr \left ( \left \{ \frac{1}{n'} d^{n'} ({U}^{n'}, y_q^{n'}) > D \right \} \right ) = 
 \Pr \left ( \left \{ \frac{1}{n'} d^{n'} ({u}^{n'}, V_q^{n'}) > D \right \} \right )
\end{align} 
This is a symmetry argument and requires the assumption of permutation invariant distortion function. The idea is that the left hand side of (\ref{JJMainDualityJJ}) depends only on the type of $y_q^{n'}$. From this it follows that the left hand side of (\ref{JJMainDualityJJ}) is equal to 
\begin{align} \label{SymmetryExpression}
\Pr \left ( \left \{ \frac{1}{n'} d^{n'} ({U}^{n'}, V_q^{n'}) > D \right \} \right )
\end{align}
where $V_q^{n'}$ is independent of $U^{n'}$. Similarly, the right hand side of (\ref{JJMainDualityJJ}) depends only on the type of $u^{n'}$ and from this it follows that the right hand side of (\ref{JJMainDualityJJ}) is also equal to (\ref{SymmetryExpression}). (\ref{JJMainDualityJJ}) follows. Details are as follows:

First step is to prove that 
\begin{align}\label{SymmetryStep1}
& \Pr \left ( \left \{ \frac{1}{n'} d^{n'} ({U}^{n'}, y_q^{n'}) > D \right \} \right ) =   \Pr \left (  \left \{ \frac{1}{n'} d^{n'} ({U}^{n'}, {y'_q}^{n'}) > D \right \} \right )
\end{align}
for sequences $y_q^{n'}$ and ${y'_q}^{n'}$ with type $q$. Since $U^{n'}$ is the uniform distribution on $\mathcal U^{n'}$, it follows that it is sufficient to prove that the sets
\begin{align}
& \left \{ u^{n'} : \frac{1}{n'} d^{n'} ({u}^{n'}, y_q^{n'}) > D \right \} \ \mbox{and} \  \left \{ u^{n'} : \frac{1}{n'} d^{n'} ({u}^{n'}, {y'_q}^{n'}) > D \right \}
\end{align}
have the same cardinality. ${y'_q}^{n'} = \pi^{n'} y_q^{n'}$ for some permutation $\pi^{n'}$ since ${y'_q}^{n'}$ and $y_q^{n'}$ have the same type. Denote the sets
\begin{align}
\mathcal B_{y_q^{n'}} \triangleq  \left \{ u^{n'} : \frac{1}{n'} d^{n'} ({u}^{n'}, y_q^{n'}) > D \right \}
\end{align}
Set $\mathcal B_{{y'_q}^{n'}}$ is defined analogously.

Let $u^{n'} \in \mathcal B_{y_q^{n'}}$. Since the distortion function is permutation invariant, $d^{n'}(\pi^{n'}u^{n'},\pi^{n'} y_q^{n'})$   $=$ $d^{n'}(u^{n'}, y_q^{n'})$. Thus, $\pi^{n'}u^{n'} \in \mathcal B_{{y'_q}^{n'}}$. If $u^{n'} \neq u'^{n'}$, $\pi^{n'}u^{n'} \neq \pi^{n'}u'^{n'}$. It follows that $|\mathcal B_{{y'_q}^{n'}}| \geq |\mathcal B_{y_q^{n'}}|$. Interchanging $y_q^{n'}$ and ${y'_q}^{n'}$ in the above argument, $|\mathcal B_{y_q^{n'}}| \geq |\mathcal B_{{y'_q}^{n'}}|$. It follows that $|\mathcal B_{{y_q}^{n'}}| = |\mathcal B_{{y'_q}^{n'}}|$. (\ref{SymmetryStep1}) follows.

Let $V_q^{n}$ be independent of $U^{n'}$. From  (\ref{SymmetryStep1}) it follows that 
\begin{align}\label{SymmetryStep1Conclusion}
& \Pr \left ( \left \{ \frac{1}{n'} d^{n'} ({U}^{n'}, y^{n'}) > D \right \} \right )  = \Pr \left ( \left \{ \frac{1}{n'} d^{n'} ({U}^{n'}, V_q^{n'}) > D \right \} \right ) 
\end{align}
By an argument identical with the one used to prove (\ref{SymmetryStep1}), it follows that 
\begin{align} \label{SymmetryStep2}
& \Pr \left ( \left \{ \frac{1}{n'} d^{n'} ({u}^{n'}, V_q^{n'}) > D \right \} \right ) =  \Pr \left ( \left \{ \frac{1}{n'} d^{n'} ({u'}^{n'}, V_q^{n'}) > D \right \} \right )
\end{align}
for $u^{n'}, u'^{n'} \in \mathcal U^{n'}$. From (\ref{SymmetryStep2}) it follows that 
\begin{align} \label{SymmetryStep2Conclusion}
& \Pr \left ( \left \{ \frac{1}{n'} d^{n'} ({u}^{n'}, V_q^{n'}) > D \right \} \right ) =  \Pr \left ( \left \{ \frac{1}{n'} d^{n'} ({U}^{n'}, V_q^{n'}) > D \right \} \right )
\end{align}
From (\ref{SymmetryStep1Conclusion}) and (\ref{SymmetryStep2Conclusion}), (\ref{JJMainDualityJJ}) follows.

\emph{Proof that a channel which is capable of communicating the uniform $X$ source with a certain distortion level is also capable of communicating bits reliably at any rate less than the infimum of the rates needed to code the uniform $X$ source with the same distortion level under the $\inf$ probability of excess distortion criterion:}

Denote 
\begin{align}
& A_{n'} \triangleq \inf_{q \in \mathcal G^{n'}}\Pr \left ( \left \{ \frac{1}{n'} d^{n'} ({U}^{n'}, y_q^{n'}) > D \right \} \right ) = 
                                         \inf_{q \in \mathcal G^{n'}} \Pr \left ( \left \{ \frac{1}{n'} d^{n'} ({u}^{n'}, V_q^{n'}) > D \right \} \right )
\end{align}
From (\ref{JCorrectCalculationJ}), it follows that rate $R$ is achievable for the channel-coding problem if 
\begin{align} \label{ChannelCalculationCriterion}
 (A_{n'}) ^{ 2^{\lfloor n' R \rfloor } - 1} \to 1 \ \mbox{as} \ n' \to \infty
\end{align}
From (\ref{JSourceErrorJ}), it follows that rate $R$ is achievable for the source-coding problem if 
\begin{align}
& (A_{n'_i})^{2^{\lfloor n'_i R \rfloor }} \to 0 \ \mbox{as} \ n'_i \to \infty  \  \mbox{for some} \ n'_i = n_0 n_i\ \mbox{for some}\  n_i \to \infty
\end{align}
Let
\begin{align} \label{JDefAlphaJ}
\alpha \triangleq \sup \{ R \ | \ (\ref{ChannelCalculationCriterion}) \ \mbox{holds} \}
\end{align}
Then, if $R' > \alpha$, 
\begin{align}\label{FinalStretch1}
& \lim_{n'_i \to \infty} (A_{n'_i})^{2^{\lfloor n_i R' \rfloor} - 1} < 1\ \forall \  R' > \alpha \  \mbox{for some sequence} \ n'_i \to \infty
\end{align}
$n'_i$ may depend on $R'$.

Then,  
\begin{align}\label{FinalStretch2}
\lim_{n'_i \to \infty} (A_{n'_i})^{2^{\lfloor n'_iR'' \rfloor} - 1}  = 0 \ \mbox{for}\ R'' > R'
\end{align}
(\ref{FinalStretch1}) and (\ref{FinalStretch2}) hold for all $R'' > R' > \alpha$. It follows that rates larger than $\alpha$ are achievable for the source-coding problem. 

Thus, a channel which is capable of communicating the uniform $X$ source with a certain distortion level is also capable of communicating bits reliably at any rate less than the infimum of the rates needed to code the uniform $X$ source with the same distortion level under the $\inf$ probability of excess distortion criterion.

\emph{Wrapping up the proof of the theorem:}

It follows that if source $U$ is directly communicated over $b$ within distortion $D$, then reliable communication can be accomplished over $b$ at rates $<R^P_U(D, \inf)$. By use of the assumption $R^P_U(D) = R^P_U(D, \inf)$, it follows that reliable communication can be accomplished over $b$ at rates $<R^P_U(D)$. In other words, the capacity of $b$, $C \geq R^P_U(D)$.
\end{proof}

\section{Discussion and recapitulation}

Randomized code constructions were made for a source-coding problem and a channel-coding problem and relation drawn between source-coding rates and channel coding-rates for the two problems. The source-coding problem is a covering problem and the channel-coding problem is a packing problem. For this reason, the connection is a randomized covering-packing connection. This duality between source-coding and channel coding is captured in (\ref{JJMainDualityJJ}).

Note {  Berger's lemma or the type covering lemma \cite{CsiszarKorner},} that at least for additive distortion functions, there exist source codes of rates approaching $R^P(D)$ such that ``balls'' around codewords cover all sequences of type $p_X$, not only a large fraction of them. Thus, in (\ref{SourceCodeD}), one does not need to take a limit; in other words, in the source-coding problem, one may not need to take a limit. Thus, a deterministic version of the source-coding problem is possible; however it is unclear, how to do the same for the channel-coding problem. For this reason, the randomized versions of the problems are needed.

The technical condition $R^P_U(D) = R^P_U(D, \inf)$ is made on the rate-distortion function. This technical condition holds for additive distortion functions, and an \emph{operational} proof which uses code constructions and various properties and relations between code constructions is provided in Chapter 5 of \cite{MukulPhdThesis}.

A proof of source-channel separation for communication with fidelity criterion follows as follows: If there exist encoder-decoder $<e^n, f^n>_1^\infty$ such that by use of this encoder-decoder, communication of source $U$ within distortion $D$ happens over a channel $k$, then, $b = <e^{n'} \circ k \circ f^{n'}>_1^\infty$ is a channel which communicates the source $U$ directly within distortion $D$. Thus, rates $<R^P_U(D)$ are achievable over $b$ by use of some encoder-decoder $<E^{n'}, F^{n'}>_1^\infty$. For this reason, reliable communication is possible over $k$ at rates $<R^P_U(D)$ by use of encoder-decoder $<E^{n'} \circ e^{n'}, f^{n'} \circ F^{n'}>_1^\infty$. By use of the standard argument of source-coding followed by channel  coding, if capacity of $k$ is $>R^P_(D)$, the uniform $X$ Source can be communicated over $k$ by source coding followed by channel coding. Proof of separation follows. The proof only uses the operational meanings of capacity (maximum rate of reliable communication) and rate-distortion function (minimum rate needed to compress a source with certain distortion),  and randomized code constructions for these problems instead of using finite-dimensional functional simplifications or finite dimensional information theoretic definitions, for example, capacity as maximum mutual information and rate-distortion function as minimum mutual information, unlike in the traditional proof of Shannon \cite{Shannon}. Functional simplifications are carried out to the extent of (\ref{JJMainDualityJJ}).

Note that whether a view or a proof is operational (in the sense used in this paper) cannot be defined mathematically precisely. However, the same can be sensed intuitively from the context in which it is used.

By use of a perturbation argument, the results can be generalized to the i.i.d. $X$ source (general $p_X$, not necessarily those for which $p_X(x)$ is rational) for additive distortion functions as discussed in Chapter 5 of  \cite{MukulPhdThesis}.

{
Finally, note that the argument to prove Theorem \ref{KKUniUniKK} uses random codes. However, if the channel is a single channel, existence of a random code implies the  existence of a deterministic code. Note further, that in the decoding rule in Theorem \ref{KKUniUniKK}, only the end-to-end description that the channel communicates the uniform $X$ source within distortion $D$ is used, and not the particular $<{b^{n'}}_1^\infty$. For this reason, even if the channel belongs to a set, that is, the channel is compound in the sense of \cite{CsiszarKorner}, Theorem \ref{KKUniUniKK} still holds. However, random codes would be needed since the argument to go from a random code to a deterministic code does not hold for a compound channel. For the same reason, a universal source channel separation theorem for communication with a fidelity criterion where universality is over the channel (channel is compound) holds if random codes are permitted. Precise details of a general, compound channel, what it means for a general, compound channel to communicate the uniform $X$ source within distortion $D$, and the capacity of a general, compound channel, are omitted.}

\bibliographystyle{IEEEtran}
\bibliography{togetherpaperBibliography.bib}

\begin{thebibliography}{1}
\providecommand{\url}[1]{#1}
\csname url@samestyle\endcsname
\providecommand{\newblock}{\relax}
\providecommand{\bibinfo}[2]{#2}
\providecommand{\BIBentrySTDinterwordspacing}{\spaceskip=0pt\relax}
\providecommand{\BIBentryALTinterwordstretchfactor}{4}
\providecommand{\BIBentryALTinterwordspacing}{\spaceskip=\fontdimen2\font plus
\BIBentryALTinterwordstretchfactor\fontdimen3\font minus
  \fontdimen4\font\relax}
\providecommand{\BIBforeignlanguage}[2]{{%
\expandafter\ifx\csname l@#1\endcsname\relax
\typeout{** WARNING: IEEEtran.bst: No hyphenation pattern has been}%
\typeout{** loaded for the language `#1'. Using the pattern for}%
\typeout{** the default language instead.}%
\else
\language=\csname l@#1\endcsname
\fi
#2}}
\providecommand{\BIBdecl}{\relax}
\BIBdecl

\bibitem{VerduHan}
S.~Verdu and T.~S. Han, ``A general formula for channel capacity,'' \emph{IEEE
  Transactions on Information Theory}, vol. 40, issue 4, pp. \ pages
  1147--1157, July 1994.

\bibitem{Shannon}
C.~E. Shannon, ``Coding theorems for a discrete source with a fidelity
  criterion,'' \emph{Institute of Radio Engineers, National Convention Record},
  vol. 7, part 4, pp. 142--163, March 1959.

\bibitem{CsiszarKorner}
I.~Csisz\'ar and J.~Korner, \emph{Information theory: coding theorems for
  discrete memoryless systems}.\hskip 1em plus 0.5em minus 0.4em\relax
  Akad\'emiai Kiad\'o, 1997.

\bibitem{Ramchandran}
S.~S. Pradhan, J.~Chou, and K.~Ramchandran, ``Duality between source and
  channel coding and its extension to the side information case,'' \emph{IEEE
  Transactions on Information Theory}, vol. 49, issue 5, pp. 1181--1203, May
  2003.

\bibitem{Ankit}
A.~Gupta and S.~Verdu, ``Operational duality between lossy compression and
  channel coding,'' \emph{IEEE Transactions on Information Theory}, vol. 57,
  issue 6, pp. 3171--3179, June 2011.

\bibitem{Yassaee}
M.~H. Yassaee, M.~R. Aref, and P.~A. Gohari, ``Achievability proof via output
  statistics of random binning,'' \emph{IEEE Transactions on Information
  Theory}, vol. 60, issue 6, pp. \ pages 6760--6786, November 2014.

\bibitem{CsiszarNarayan}
I.~Csiszar and P.~Narayan, ``Channel capacity for a given decoding metric,''
  \emph{IEEE Transactions on Information Theory}, vol. 41, issue 1, pp. \ pages
  35--43, January 1995.

\bibitem{MukulPhdThesis}
M.~Agarwal, ``A universal, operational theory of multi-user communication with
  fidelity criteria,'' Ph.D. dissertation, Massachusetts Institute of
  Technology, February 2012.

\bibitem{ShannonReliable}
C.~E. Shannon, ``A mathematical theory of communication,'' \emph{Bell System
  Technical Journal}, vol.~27, pp. 379--423 (Part 1) and pp. 623--656 (Part 2),
  July (Part 1) and October (Part 2) 1948.

\end{thebibliography}

\appendix

\section{Intuitive explanation of the randomized covering-packing duality} \label{AppendixIntuitive}

This appendix explains on an intuitive level, the covering-packing duality. The authors emphasize that mathematically this section is imprecise and is only for the purpose of developing intuition.

A general channel which directly communicates the uniform $X$ source to within a distortion $D$ can be intuitively thought of as follows: with high probability, a sequence in $\mathcal U^{n'}$ is communicated with distortion $\leq nD'$ and this probability $\to 0$ as $n' \to \infty$. See Figure \ref{intuitiveChannel} in Section \ref{Introduction} with $n$ replaced with $n'$.

\begin{figure} 
\begin{center}
\includegraphics[scale = 0.8]{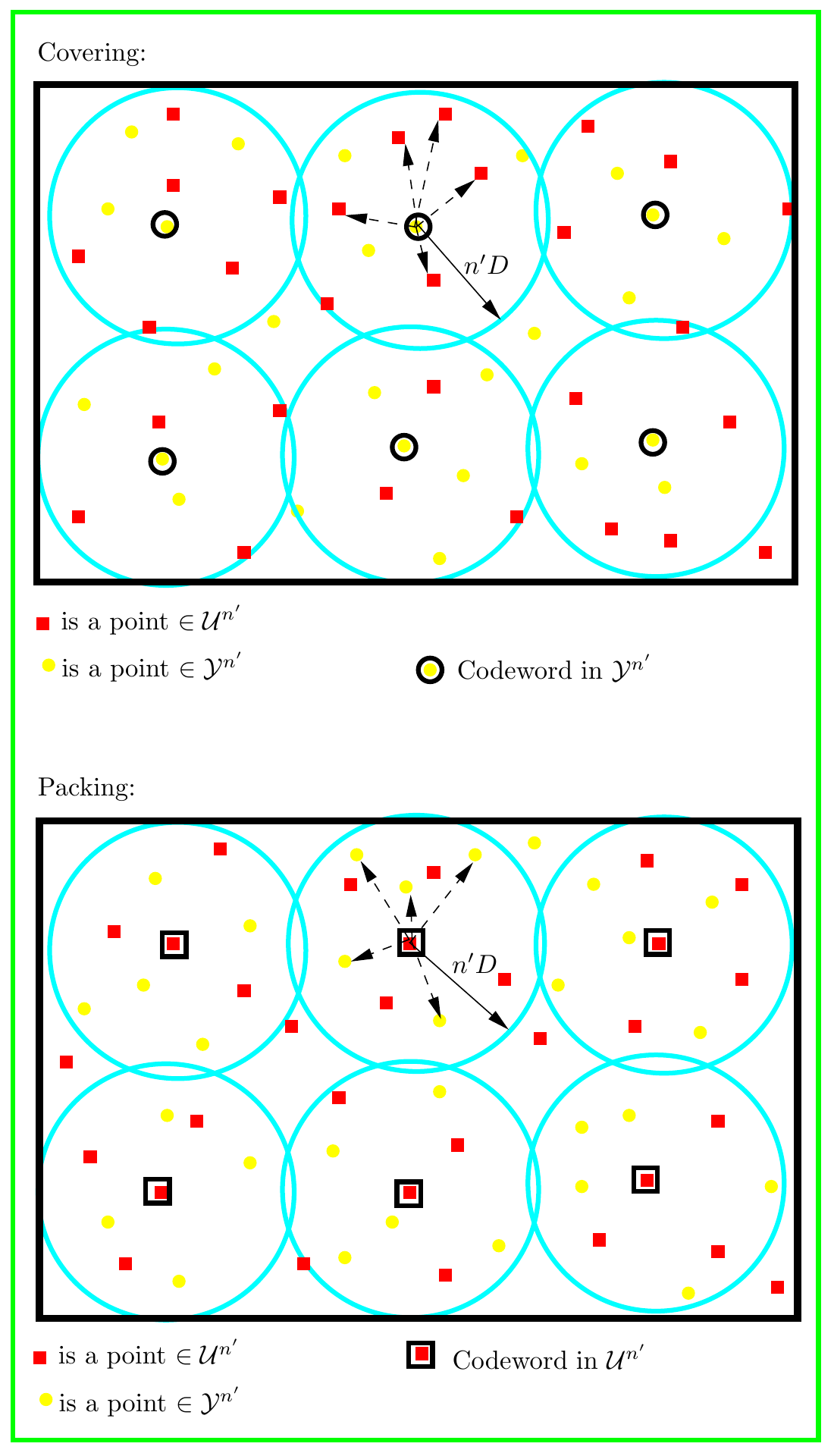}
\caption{Covering: what is the minimum number of balls (equivalently, the number of codewords $\in \mathcal Y^{n'}$) with centers around certain $y^{n'} \in \mathcal Y^{n'}$ and balls in $\mathcal U^{n'}$ which cover the whole $\mathcal U^{n'}$ space. Packing: what is the maximum number of balls with (equivalently, the number of codewords $\in \mathcal U^{n'}$)  centers around certain $u^{n'} \in \mathcal U^{n'}$ such that these balls pack the $\mathcal Y^{n'}$ space. Note that balls in the covering problem have centers $\in \mathcal Y^{n'}$ but the balls are in $\mathcal U^{n'}$ whereas balls in the packing problem have centers $\in \mathcal U^{n'}$ but the balls are in $\mathcal Y^{n'}$}
\label{deterministicCP}
\end{center}
\end{figure}

The \emph{deterministic} (as opposed to randomized) covering-packing, or the source coding-channel coding problem in our setting, on an intuitive level is pictured in Figure \ref{deterministicCP}. For the covering problem, with reference to this figure, $y^{n'} \in \mathcal Y^{n'}$ but balls of radius $n'D$ around $y^{n'}$ are made in the $\mathcal U^{n'}$ space. The source-coding question is: what is the minimum number of balls of radius $n'D$ which cover the $\mathcal U^{n'}$ space. In other words, what is the minimum number of $y^{n'} \in \mathcal Y^{n'}$ such that balls of radius $nD$ around these $y^{n'} \in \mathcal Y^{n'}$ cover the $\mathcal U^{n'}$ space. For the packing problem, first recall the intuitive action of the channel depicted in Figure \ref{intuitiveChannel}. With reference to the figure, for the packing problem, the question is to find the minimum number of what is the maximum number of $u^{n'} \in \mathcal U^{n'}$ such that balls of radius $n'D$ around these $u^{n'}$ pack the $\mathcal Y^{n'}$ space. 

\begin{figure} 
\begin{center}
\includegraphics[scale = 1.0]{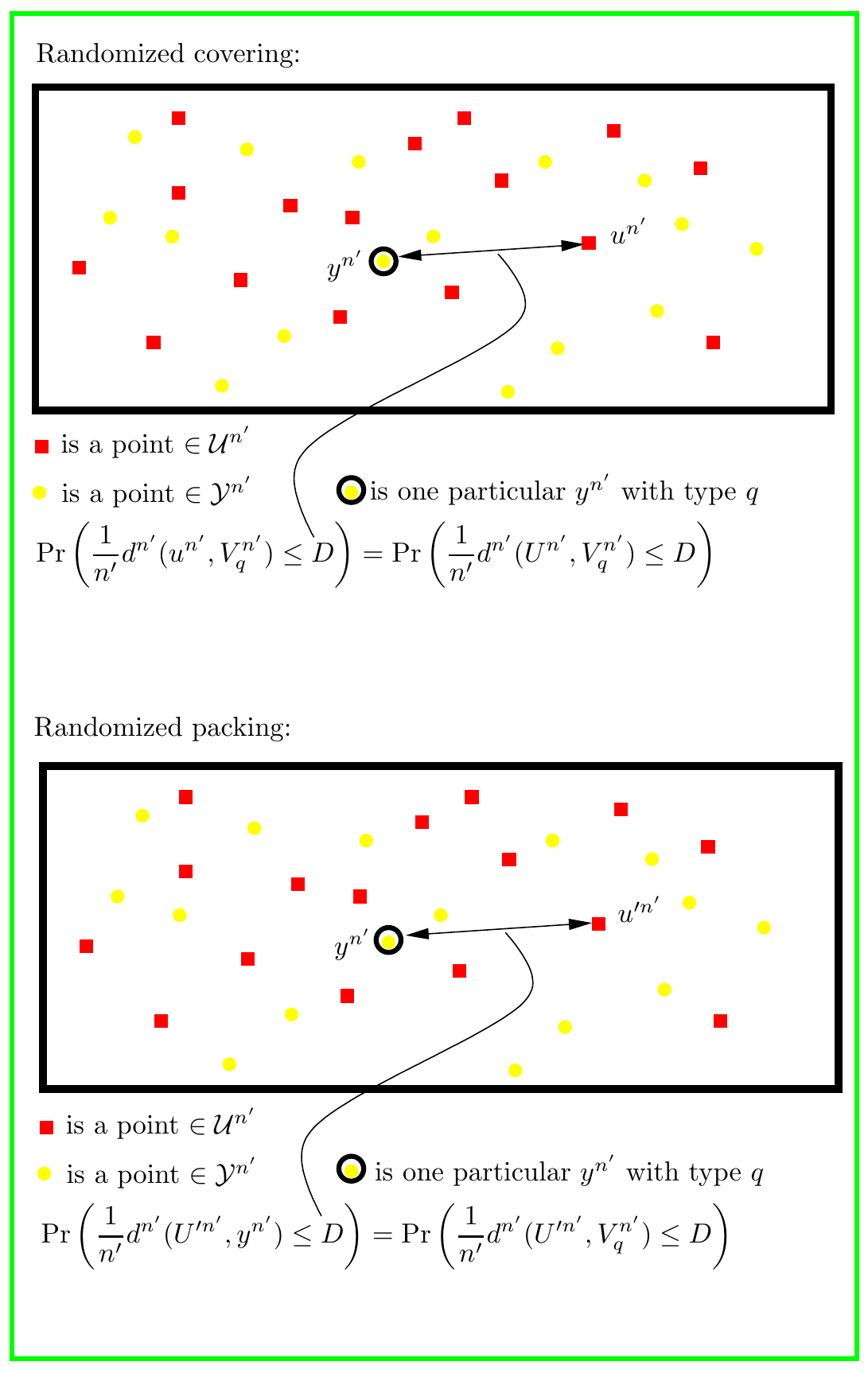}
\caption{The randomized covering-packing picture for the problem of communication with a fidelity criterion}
\label{randomizedCP}
\end{center}
\end{figure}

The randomized covering-packing picture is figuratively described in Figure \ref{randomizedCP}. 

In the covering problem, let the block-length be $n'$.  Suppose $u^{n'}$ needs to be compressed. Suppose a  codeword of type precisely $q$ is generated uniformly from the set of all sequences with type precisely $q$. Denote this uniform distribution by $V_q^{n'}$ and a realization of $V_q^{n'}$ by $y^{n'}$. Probability that $y^{n'}$ will code $U^{n'}$ is 
\begin{align}
\Pr \left ( \frac{1}{n'} d^{n'}(u^{n'}, V_q^{n'}) \leq D\right ) 
\end{align}
This probability is independent of $u^{n'}$ by symmetry because the distortion metric is 
\begin{align}
\Pr \left ( \frac{1}{n'} d^{n'}(U^{n'}, V_q^{n'}) \leq D\right ) 
\end{align}
The way things \emph{intuitively} work for increasing block-lengths, the number of sequences needed to code the source $U^{n'}$ if codewords of type $q$ are used is approximately 
\begin{align}
\frac{1}{\Pr \left ( \frac{1}{n'} d^{n'}(U^{n'}, V_q^{n'}) \leq D\right ) }
\end{align}.

$q$ is arbitrary and thus, with this coding scheme, the number of codewords to code the uniform $X$ source is approximately
\begin{align}
\inf_{q} \frac{1}{\Pr \left ( \frac{1}{n'} d^{n'}(U^{n'}, V_q^{n'} ) \leq D\right )}  \triangleq \beta
\end{align}
In general, there may be a scheme for which number of codewords needed is $\leq \beta$.

In the packing problem, generate $2^{n'R}$ codewords independently and uniformly from $\mathcal U^{n'}$. Suppose $u^{n'}$ is transmitted. By the action of the channel, it follows that with high probability, $y^{n'}$ is received such that 
\begin{align}
\frac{1}{n'} d^{n'}(u^{n'}, y^{n'}) \leq D
\end{align}
Let the type of the received sequence $y^{n'}$ be $q$. Let $u^{n'}$ be another non-transmitted codeword which is generated using $U'^{n'}$. Note that $U^{n'}$ and $U'^{n'}$ are the same in distribution. Probability that there might be a mistake to say that $u'^{n'}$ is transmitted is 
\begin{align}
\frac{1}{n'} d^{n'}(U'^{n'}, y^{n'}) \leq D
\end{align}
The above probability is the same for all $y^{n'}$ by symmetry because the distortion metric is permutation invariant, and hence, is equal to 
\begin{align}
\Pr \left ( \frac{1}{n'} d^{n'}(U'^{n'}, V_q^{n'}) \leq D \right )
\end{align}
where $V_q^{n'}$ is defined in the above discussion on covering. Note that $q$ is arbitrary and in order to get a bound on the total number of allowed codewords, the worst possible $q$ needs to be considered.  The way union bound works and the way things work for large block-lengths, the number of sequences which can be chosen as codewords for the channel-coding problem is
\begin{align}
\inf_{q}\frac{1}{\Pr \left ( \frac{1}{n'} d^{n'}(U'^{n'}, V_q^{n'}) \leq D \right )} = \beta
\end{align}
In general, there may be a scheme for which number of codewords  is $\geq \beta$.

Finally, note that the $\beta$ in the covering and packing problem are the same. It follows that $C \geq R^P_U(D)$ where $C$ is the capacity of the channel.

This is the intuitive base behind the proof of Theorem \ref{KKUniUniKK} and the resulting duality. Note further that this section is only for the sake of intuition and is mathematically imprecise. Precise proof have been provided in the proof of Theorem \ref{KKUniUniKK}.

\end{document}